\documentclass[aps,pre,showpacs,floats,twocolumn,superscriptaddress,floatfix,footinbib]{revtex4-1}
\usepackage{graphicx}
\usepackage{amsmath,amsthm,amsfonts,amssymb,bm}
\usepackage{wasysym}
\usepackage{hyperref}
\usepackage[usenames,dvipsnames]{color}

\graphicspath{{../Figs/}}

\newcommand{\beq}{\begin{equation}}
\newcommand{\eeq}{\end{equation}}
\newcommand{\beqa}{\begin{eqnarray}}
\newcommand{\eeqa}{\end{eqnarray}}
\newcommand{\beqan}{\begin{eqnarray*}}
\newcommand{\eenan}{\end{eqnarray*}}

\def\Bbb{\mathbb}

\newcommand{\Dslash}{{\slash{\kern -0.5em}\partial}}
\newcommand{\Aslash}{{\slash{\kern -0.5em}A}}

\def\sqr#1#2{{\vcenter{\hrule height.#2pt
     \hbox{\vrule width.#2pt height#1pt \kern#1pt
        \vrule width.#2pt}
     \hrule height.#2pt}}}

\def\square{\mathchoice\sqr68\sqr68\sqr{4.2}6\sqr{3.0}6}

\def\thinspace{\kern .16667em}

\def\xp{x_{{\kern -.2em}_\perp}}
\def\subp{_{{\kern -.2em}_\perp}}

\newcommand{\eqdef}{ {\kern 0.2em}={\kern -0.5em}:{\kern 0.2 em} }
\newcommand{\defeq}{ {\kern 0.2em}:{\kern -0.5em}={\kern 0.2 em} }

\newcommand{\Nat}{{\mathbb N}}
\newcommand{\Int}{{\mathbb Z}}

\newcommand{\downtri}{\raisebox{1pt}{$\bigtriangledown$}}
\newcommand{\uptri}{{$\bigtriangleup$}}
\newcommand{\BZ}{\mathrm{BZ}}
\newcommand{\xfer}{{\Bbb T}}

\newcommand{\Num}{{\cal N}}

\newcommand{\Vac}{{\varnothing}}

\begin{document}

\title{Communicating through a geometrically frustrated channel}
 \author{Amir Nourhani}
 \email{Amir.Nourhani@gmail.com}
 \affiliation{Department of Physics, Pennsylvania State University, University Park, PA 16802}
 \author{Vincent H. Crespi}
 \affiliation{Department of Physics, Pennsylvania State University, University Park, PA 16802}
 \affiliation{Department of Materials Science and Engineering, Pennsylvania State University, University Park, PA 16802}
 \affiliation{Department of Chemistry, Pennsylvania State University, University Park, PA 16802}
 \author{Paul E. Lammert}
 \email{lammert@psu.edu}
 \affiliation{Department of Physics, Pennsylvania State University, University Park, PA 16802}

\begin{abstract}
We propose an intuitively appealing formulation of the zero-temperature triangular lattice Ising antiferromagnet (TIAFM) on a cylinder as a model of noninteracting fermions hopping on a ring and evolving in imaginary time with pair annihilation events. Among the features of the model which can then be related to local semi-conservation of particle number are: infinite-range influence of boundary conditions, multiple ``zero temperature pure phases'' in the infinite-length limit differing in entropy density, sensitivity of the asymptotic rate of decay (with respect to length) of mutual information between end configurations to circumference modulo 3, and even the known power-law falloff of the spin-spin correlator on an infinite plane. The ability of boundary conditions to determine the bulk entropy density enables communication between the two far-separated ends of such a cylinder even when all pair-wise spin-spin correlations between the ends are vanishingly small. In the fermionic language, breakdown of these zero-temperature phenomena at positive temperature is understood as passage of the system into a superconducting phase.
\end{abstract}


\maketitle 

An archetype of {\em geometrical frustration} --- incompatibility of elementary interactions --- is the antiferromagnetic Ising model on a triangular lattice (TIAFM), each elementary triangle of which can contain at most two satsified bonds. The result, even at zero temperature, is a disordered macrostate with a nonzero entropy density. How different is this frustration-induced disorder from thermal disorder? It has long been known\cite{Stephenson-64,Stephenson-70} that the spin-spin correlation length of the zero-temperature TIAFM is infinite, so disorder is subtly weaker than it might appear at first.

\begin{figure}\centering
\includegraphics[width=90mm]{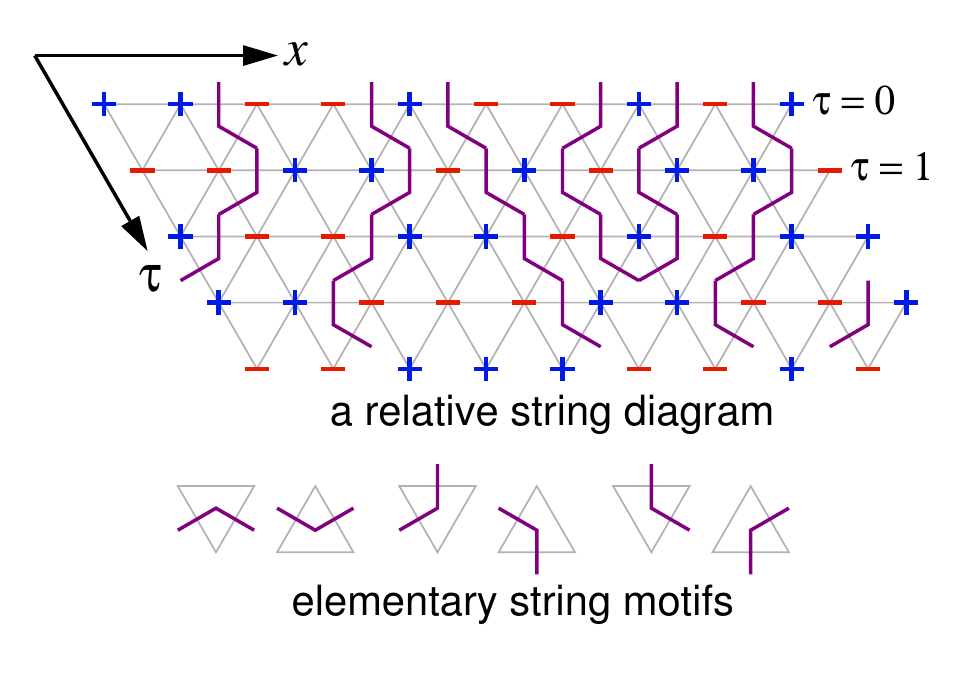}
\caption{
When each bond of a finite lattice belongs to a unique down-pointing triangle (\downtri), a ground microstate must have exactly one unsatisfied bond in each \downtri, but may have one or three in a \uptri. Taking the ground microstate with all spins up/down on even/odd rows as a reference, any microstate (not only ground) can be indicated up to global spin flip  by the boundaries (purple ``strings'') between the regions matching the  reference microstate and its global spin flip. The strings in the string diagram cannot terminate inside the system. Ground spin microstates are in two-to-one correspondence with string diagrams using any of the displayed motifs except the left-most one. If strings are conceived of as particle worldlines, a local semi-conservation law emerges, with the second motif corresponding to pair annihilation.
\label{fig:fiducial}}  
\end{figure}

Although it is a commonplace that reducing dimensionality suppresses ordering, we show here that dropping the dimensionality of the TIAFM by one, to a cylindrical geometry, {\em accentuates} the subtle order of the TIAFM ground state sufficiently that the two ends of a very long cylinder can actually communicate at long range through the sea of frustration-induced disorder. Consider a cylindrical TIAFM system of fixed circumference, but variable length, formed by identifying the left and right edges of a planar system such as shown in Fig.~\ref{fig:fiducial}. One can regard such a system as a chain of rings of spins. From this perspective, elementary degrees of freedom are configurations of entire rings. Mutual information~\cite{Shannon+Weaver, Billingsley-ETI, Csiszar+Korner, Cover+Thomas} provides a natural tool to study correlations of such complex degrees of freedom. One expects that the mutual information between the ends (top and bottom) of such a cylindrical system will not only fall off exponentially, as for a thermally disordered system with short-range interactions, but with a decay rate varying smoothly and monotonically with circumference. However, we have found that the mutual information decay rate is {\em not} monotonic in circumference ($L$), but sensitive to $L$ mod 3, as well as to whether the cylinder is wrapped with periodic or antiperiodic boundary conditions around the circumference. Moreover, for antiperiodic wrapping and $L$ a multiple of 3, the decay is even subexponential (inverse square of the length). In this Letter, we show an equivalence of the zero-temperature cylindrical TIAFM with a model of fermions hopping on a ring which allows a simple and unified explanation of this puzzling behavior, along with the infinite-range influence of boundary conditions and power-law falloff of spin-spin correlations in the planar limit. The fermions --- corresponding to satisfied circumferential bonds --- are noninteracting, but pairs of neighboring particles can annihilate during evolution in imaginary time (downward along the length of the cylinder in Fig.~\ref{fig:fiducial}). This {\em semi-conservation} is key to the behavior of the end-to-end mutual information. It also implies the capacity of boundary conditions at the cylinder ends to have an infinite-range influence, resulting in multiple ``zero-temperature pure phases'' in an infinitely long cylindrical system, labelled by the number of satisfied circumferential bonds and differing in entropy density. In principle, this enables long-range communication {\em through} the disordered ground state by using boundary conditions at one end to control the entropy density observed at the other. In the planar limit, local particle-number semi-conservation, effectively promoted to full conservation, explains even the power-law falloff of the spin-spin correlation function. All these aspects are explained below.

At positive temperature, all of these unusual phenomena give way to 
more ordinary behavior, i.e., exponential falloff and a unique phase,
for large enough systems. In fermionic language, this means that
there is a nonzero excitation gap even in the large-system limit, while
simultaneously particle-number fluctuations strengthen qualitatively.
The way to achieve that is to become superconducting. We 
indicate the relation of the superconducting order parameter 
$\langle c_x c_{x+1}\rangle$ to the density of ``excited'' triangles 
with certain kinds of near-neighbor environments.

To reach the fermion formulation, we first think in terms of bond configurations
rather than spin configurations, represent bond configurations as {\it relative string
diagrams\/}, interpret the strings as fermion worldlines, and finally deduce
an appropriate transfer matrix or Hamiltonian written in terms of creation and
annihilation operators.
Geometrically, our systems are of the sort depicted in Fig.~\ref{fig:fiducial},
hence decompose into down-pointing triangles (\downtri's). Thus, the ground
microstate condition is two satisfied bonds in each \downtri. There is no
such condition for up-pointing triangles (\uptri's). They may have either 
no or two satisfied bonds. To represent configurations by strings, begin
with a reference configuration containing alternating horizontal rows of 
up and down spins so that all horizontal bonds are unsatisfied.
Crossing each satisfied bond with a line produces strings of the
motifs \includegraphics[width=18pt]{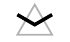} and 
\includegraphics[width=18pt]{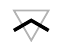} running across the system.
Now, if for some other ground microstate a similar diagram is constructed,
using the full palette of six motifs shown in Fig.~\ref{fig:fiducial},
and then a Boolean XOR operation taken with the reference diagram
(that is, keep string links which occur in one or the other diagram, but
not both), the result is the relative string diagram for the microstate in question.
An example is shown in Fig.~\ref{fig:fiducial}. One can easily see that such
a diagram can never have free ends inside the system. The motif
\includegraphics[width=18pt]{motif6} cannot occur, but 
\includegraphics[width=18pt]{motif5} can (as in Fig.~\ref{fig:fiducial}).
This suggests regarding the strings as representing particle worldlines:
time increasing from top to bottom in the diagram, and the motif 
\includegraphics[width=18pt]{motif5} 
corresponds to annihilation of pairs of neighboring particles.
One such event is depicted in Fig.~\ref{fig:fiducial}(a).
The particle number $\Num$ can therefore change only by pair annihilation
(decreasing it by 2), or particles entering or leaving at the boundary.
On a cylinder, obtained by identifying the left and right edges of a
parallelogram as in Fig.~\ref{fig:fiducial}a 
(with periodic or antiperiodic spin boundary conditions),
there is no boundary, resulting in a local and global 
semi-conservation law for particle number. 
Henceforth, we will be concerned only with cylindrical geometries,
possibly in the limit that ring diameter $L$ goes to infinity. 

\begin{figure}\centering
  \begin{picture}(220,150)
\put(35,0){ fraction of satisfied horizontal bonds \large $n$}
\put(-15,45){\rotatebox{90}{entropy density \large $\sigma$}}
\put(0,10){\includegraphics[width=80mm]{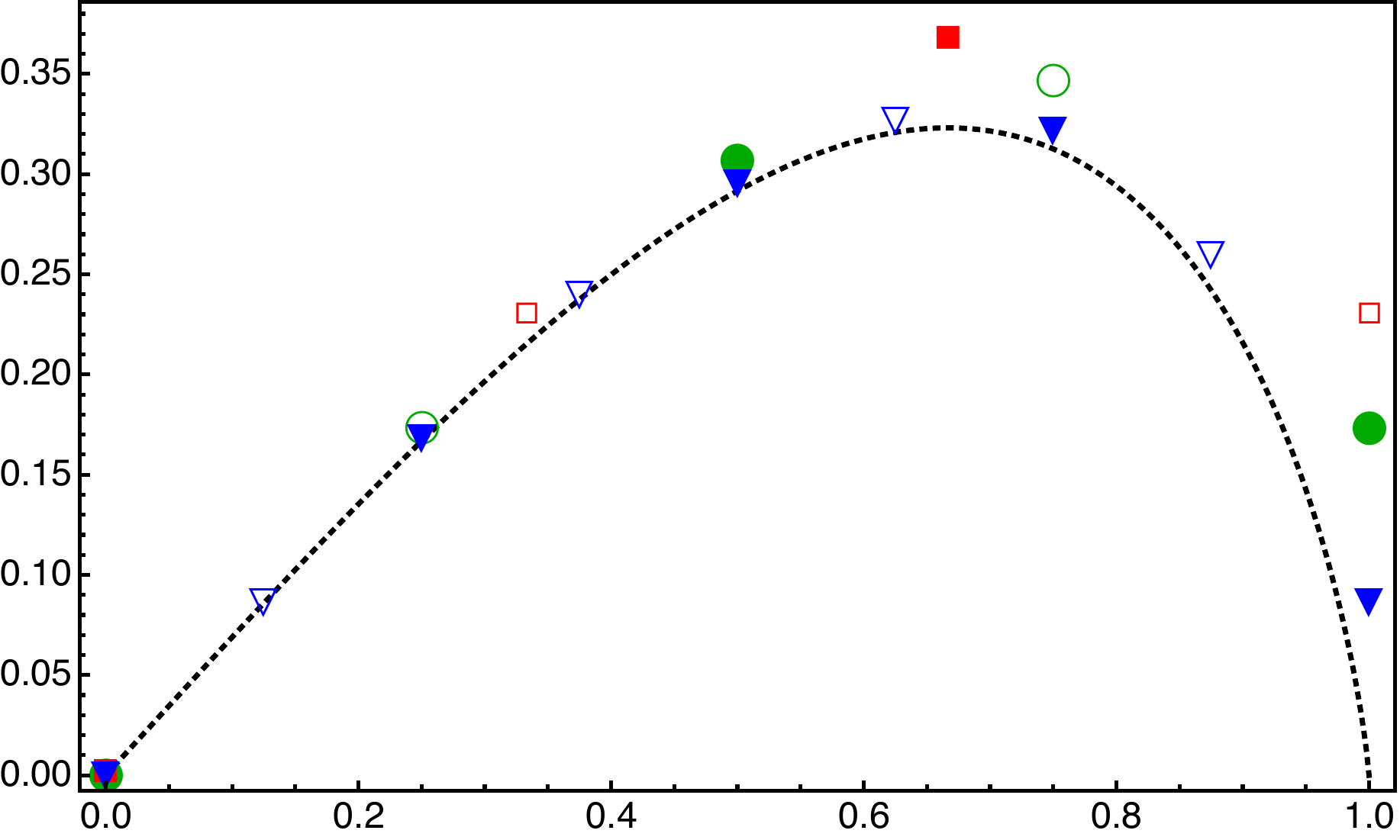}}
  \end{picture}
\caption{Entropy density versus fraction of unsatisfied circumferential
bonds (equivalently, 1 minus the fermion density) for the zero-temperature
TIAFM on a cylinder of circumference $L$. 
The solid curve is from Eq.~(\ref{eq:entropy-density}).
Symbols: exact numerical results, with solid (open) symbols denoting periodic (antiperiodic)
boundary conditions, and for circumference $L$ equal to 3 ($\square$), 
4 ($\diamondsuit$) and 8 ($\triangledown$).
With the exception of $n=1$, the infinite-$L$ limit is nearly achieved already for
$L=8$. At $n=1$, the entropy density is exactly $(\ln 2)/L$.
\label{fig:master-curve}}  
\end{figure}
While the string representation provides much qualitative insight, 
a fermionic representation allows quantitative calculations; 
Fermi statistics will take care of the nonintersecting nature of
the strings without need for a particle interaction.
Taking each horizontal bond position as a fermionic degree of freedom
[(un)satisfied=(un)occupied], and observing the possibilities of string
movement from time $\tau$ to $\tau+1$, the transfer matrix with the
fifth (pair annihilation) motif held in abeyance is seen to satisfy
the conditions
\begin{equation}
\xfer_0 c_x^\dagger = 
(c_{x-1}^\dagger + c_{x}^\dagger) \xfer_0, \quad \xfer_0 |\Vac\rangle 
= |\Vac\rangle.
\label{eq:T0-conditions-x}
\end{equation}
By working in momentum space, one easily deduces the form
\begin{equation}
\xfer_0 = e^{-{H}_0 + iP/2},  
\label{eq:T-exponential-form}
\end{equation}
where
\begin{equation}
{H}_0 = \sum_{q\in \BZ} {\varepsilon}(q) n(q),
\qquad
P = \sum_{q\in\BZ} q\, n(q),
\label{eq:H0}
\end{equation}
can be considered a Hamiltonian and total momentum operator, respectively,
with mode occupation operators $n(k) = c(k)^\dagger c(k)$,
and effective mode energies
\begin{equation}
{\varepsilon}(q) = -\ln\left(2\cos\frac{q}{2}\right).
\label{eq:mode-energy}
\end{equation}
Since, ignoring $P$ in the exponent (``physical'' states have zero net momentum, anyway),
the transfer matrix (\ref{eq:T-exponential-form}) has the form of 
a quantum evolution operator through one unit of imaginary time, this is an example
of the ubiquitous correspondence\cite{Roepstorff} between imaginary-time quantum evolution and
classical statistical systems such as is exploited in diffusion quantum 
Monte Carlo\cite{Foulkes+01}.

The allowed modes depend on the parity of the number of particles $\Num$ 
according to
\begin{equation}
\label{eq:BZ}
\BZ = 
\begin{cases}
\frac{2\pi}{L}\Int\cap (-\pi,\pi], & \Num \, \mathrm{odd}
\\
\frac{2\pi}{L}\left(\Int+\frac{1}{2}\right)\cap (-\pi,\pi], & \Num\, \mathrm{even}.
\end{cases}
\end{equation}
Since an even number of spin reversals must occur on circling the ring, 
periodic or antiperiodic boundary conditions are dictated by $L$ and the
parity of $\Num$.

Reinstating the fifth motif, the complete zero-temperature transfer matrix
can be written in a factorized form as
\begin{equation}
\label{eq:T-two-exps}
\xfer = e^{-H_0} e^{-H_{\mathrm{pr}}},
\end{equation}
where
\begin{equation}
\label{eq:T-pair-destruction}
H_{\mathrm{pr}} = -\sum_{i\in{\Bbb L}} c_i c_{i+1} = \sum_{0 < q\in\BZ} 2i(\sin q)\,  c(-q) c(q).
\end{equation}

Consider now cylinders of circumference $L$ and length $T$.
With periodic boundary conditions along the length, but restricted to
fixed particle number $\Num$ (density $n=\Num/L$),
we easily compute the entropy density $\sigma$:
\begin{equation}
e^{L T \sigma} 
= \mathrm{Tr}_\Num \xfer_0^T 
= \sum_{E\in \,\mathrm{spec}\, H_0} e^{-E T}  \sim e^{-E_0(n) T},
\end{equation}
where the final expression is the $T\to\infty$ asymptote and
$E_0(n)$ is the energy of a Fermi sea.
For large $L$ and at Fermi momentum $k_F = \pi n$,
\begin{equation}
\sigma = -\frac{E_0}{L} = -\int_0^{k_F} \varepsilon(q)\, \frac{dq}{\pi}.
\label{eq:entropy-density}
\end{equation}
This entropy density (previously obtained by 
Dhar {\it et al.}\cite{Dhar+Chaudhuri+Dasgupta-00} in a way we
feel is much less intutive)
is plotted in Fig.~\ref{fig:master-curve}, along with exact 
calculations for small circumferences demonstrating that the limit is
reached very quickly with $L$.
That result makes clear that for more general boundary conditions, 
with limiting satisfied-bond density $n(-\infty)$ at the top 
not less than that at the bottom, $n(+\infty)$,
the particle density in the bulk will be the unique value in 
$[n(\infty),n(-\infty)]$ maximizing $\sigma(n)$, because the
number of jumps of particle number is limited by half the circumference.
Hence, as the length goes to infinity, the semi-conservation of particle
number is effectively promoted to a full conservation law.
The corresponding macrostates can be considered zero-temperature
pure phases. For an infinite cylinder (ring indices in $\Int$) 
in the zero-temperature pure phase labelled by particle number $\Num$, 
the joint probability of configurations $X_0$ on ring $0$ and 
$X_\tau$ on ring $\tau$ is
\begin{equation}
P_\Num(X_0=a \,\&\, X_\tau=b) = 
          \langle \varphi_{\Num,0}|b \rangle 
          \frac{Z_\tau(b|a)}{\lambda_{\Num,0}^\tau}
          \langle a|\varphi_{\Num,0}\rangle,
\label{eq:P-infinite-cylinder}
\end{equation}
where $\lambda_{\Num,0} = e^{-E_{\Num,0}}$ is the largest eigenvalue of
the transfer matrix $\xfer_0$ in the $\Num$-particle subspace.

We turn now to the 
{\it mutual information}~\cite{Shannon+Weaver,Billingsley-ETI,Csiszar+Korner,Cover+Thomas} 
between the spin configurations on distinct rings.
Recall that the mutual information $I(X\! :\! Y)$ between discrete random 
variables $X$ and $Y$ can be expressed as
  \begin{equation}
\label{eq:mutual-info}
I(X\! :\! Y) = H(Y) - H(Y|X),
  \end{equation}
where 
$H(Y|X) = -\sum_x P_X(x) \sum_y P_{Y|X}(y|x)\ln P_{Y|X}(y|x)$ is the conditional
entropy of $Y$ given $X$ and $H(Y)$ is the (unconditional) entropy of $Y$.
Mutual information is of increasing interest in classical statistical
mechanics\cite{Lau+Grassberger-13,Wilms+-11,Melchert+Hartmann-13}, as well
as quantum information theory\cite{Nielsen+Chuang-Book}.

Starting from (\ref{eq:P-infinite-cylinder}), one shows
that the the mutual information between ring configurations $X_0$ and $X_\tau$
is, asymptotically in $\tau$, 
\begin{equation}
\label{eq:infinite-cylinder-MI}
I_{\infty,\Num}(X_0\! :\! X_\tau) 
\sim \left| \frac{\lambda_{\Num,1}}{\lambda_{\Num,0}} \right|^{2\tau}.
\end{equation}
Where $\lambda_{\Num,1}$ is the subleading eigenvalue of $\xfer_0$
in the $\Num$-particle subspace (it corresponds to a particle-hole
excitation and is not unique, but the modulus is).
It is remarkable that the coefficient is exactly one and
the falloff rate is twice that of an ordinary correlation function,
but the dependence on $L$ is simple monotonic decrease of the decay rate.
The exponent is easily explained.
Writing
\begin{equation}
I_{\infty,\Num}(X_0\! :\! X_\tau) 
= \sum_i P_{X_0}(i) \left( H(X_\tau) - H(X_\tau|X_0=i) \right),
\nonumber
\end{equation}
and expanding in powers of the deviations 
$\delta P(j|i) = P_{X_\tau}(j|X_0=i) - P_{X_\tau}(j)$
of the conditional probabilies from the marginals, one finds that the
first order terms cancel.
Hence, $I_{\infty,\Num}(X_0\! :\! X_\tau)$
is second order in the deviations, which are themselves essentially connected
correlation functions with a decay rate equal to the energy gap. 

\begin{figure}
\setlength{\unitlength}{1cm}
\begin{picture}(10,7)
\put(-0.2,-0.2){\includegraphics[width=80mm]{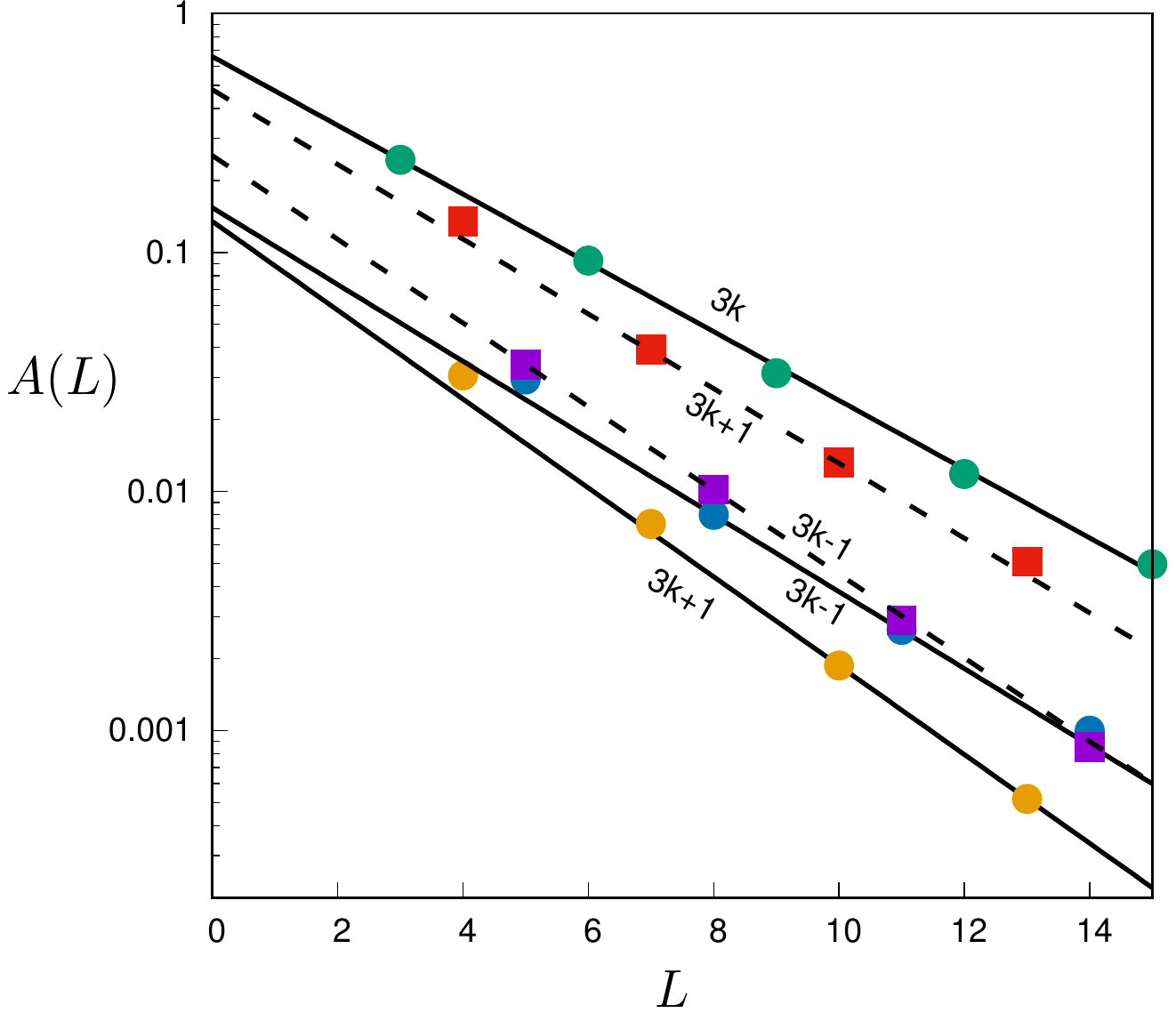}}
\put(1.8,1.0){$\blacksquare$ : odd $\Num$}
\put(1.8,1.5){\CIRCLE : even $\Num$}
\end{picture}
\caption{Amplitudes of the asymptotic decay of end-to-end mutual 
information break into families labelled by $\Num$ parity and residue 
class $L$ mod 3 ($3k+m$, $m=0,\pm 1$). 
Within each family, $A(L)$ decays approximately exponentially in $L$.
}
\label{fig:mi-amplitudes}  
\end{figure}
Mutual information between the end configurations of a finite cylinder
with free boundary conditions has a much richer structure.
The relevant spectral gap is that above the minimum energy
state ($n \approx 2/3$, see Fig.~\ref{fig:master-curve}) in the 
{\em full\/} configuration space. 
A simple calculation reveals that these are never of particle-hole
type, but rather particle-particle (pp) or hole-hole (hh), 
as tabulated in Table \ref{tab:energy-gaps}. 
\begin{table}[h]
\setlength{\tabcolsep}{6pt}
  \centering
\begin{tabular}{c|ccc|ccc}
\multicolumn{1}{c|}{$\Num$ parity} & \multicolumn{3}{c|}{ even }  & \multicolumn{3}{c}{ odd } \\
\hline
$L$ mod 3                          & 0      & 1    & 2    & 0   & 1    & 2 \\
excitation type                    & hh     & pp   &  hh  &  hh & hh   & pp  \\
$\Delta\varepsilon\cdot L/(\pi\sqrt{3})$  &  1     & 1/3  & 1/3  & 0   & 2/3  & 2/3 
\end{tabular}
\caption{
\label{tab:energy-gaps}
Fundamental energy gaps $\Delta\varepsilon$ of $\xfer_0$ and corresponding excitation types. 
`hh' and `pp' indicate excitations involving removal (addition) of two particles.
Energies are reported in units of 
$v(q_0)\delta = (\sqrt{3}/2)(2\pi/L) = \pi\sqrt{3}/L$, 
according to a linearized approximation of $\varepsilon(q)$. 
Due to the strict convexity of $\varepsilon(q)$, energies reported for
hh (pp) excitations are overestimates (underestimates), though
the relative error goes to zero as $L\to\infty$. 
}
\end{table}
\begin{figure}
{\includegraphics[width=80mm]{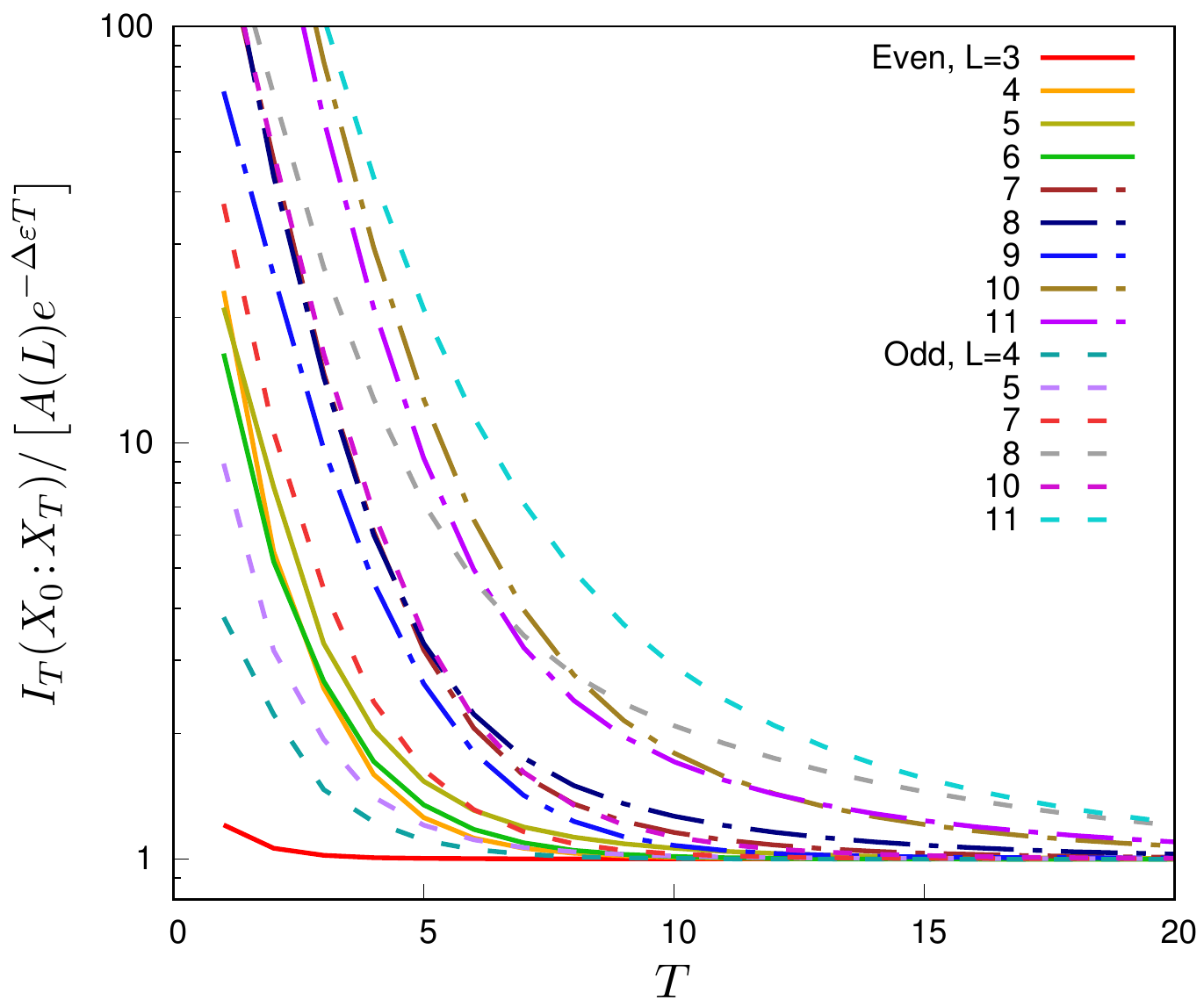}}
\caption{Ratio of the end-to-end mutual information 
\hbox{$I_T(X_0\! :\! X_T)$} to the leading behavior 
$A(L) e^{-\Delta\varepsilon T}$, for cases without zero-energy modes
and $3 \le L \le 11$. The approach is always from above because
the leading behavior correctly accounts for all information transmission
via the first two eigenmodes, but not the others.
}
\label{fig:end-to-end-mi}  
\end{figure}

Asymptotically in the length $T$ of the cylinder, the end-to-end
mutual information behaves as
\begin{equation}
\label{eq:end-to-end-mi-nonexceptional}
I_T(X_0\! :\! X_T) \sim 
A(L) \left(\frac{\lambda_1}{\lambda_0}\right)^T.
\end{equation}
For small values of $L$,
$A(L)$ is plotted in Fig.~\ref{fig:mi-amplitudes}
and the ratio of the exact mutual information to the asymptotic 
formula in Fig.~\ref{fig:end-to-end-mi}.

Not only does the end-to-end mutual information depend on 
different spectral gaps than ring-to-ring mutual information in the bulk, 
but the decay rate is precisely the energy gap, rather than twice it. 
These features are related.
Fluctuations in an infinite cylinder are typical of the phase
it is in, but correlation between all fluctuations on distinct
rings decay with distance.
In contrast, the fluctuations which dominate the
end-to-end mutual information can be described as global
fluctuations of the entire system between phases.  
As a result, the expansion in powers of the deviations 
described above is invalid because the relevant deviations are 
comparable to the unconditioned probabilities.

\begin{figure}
\includegraphics[width=8.5cm]{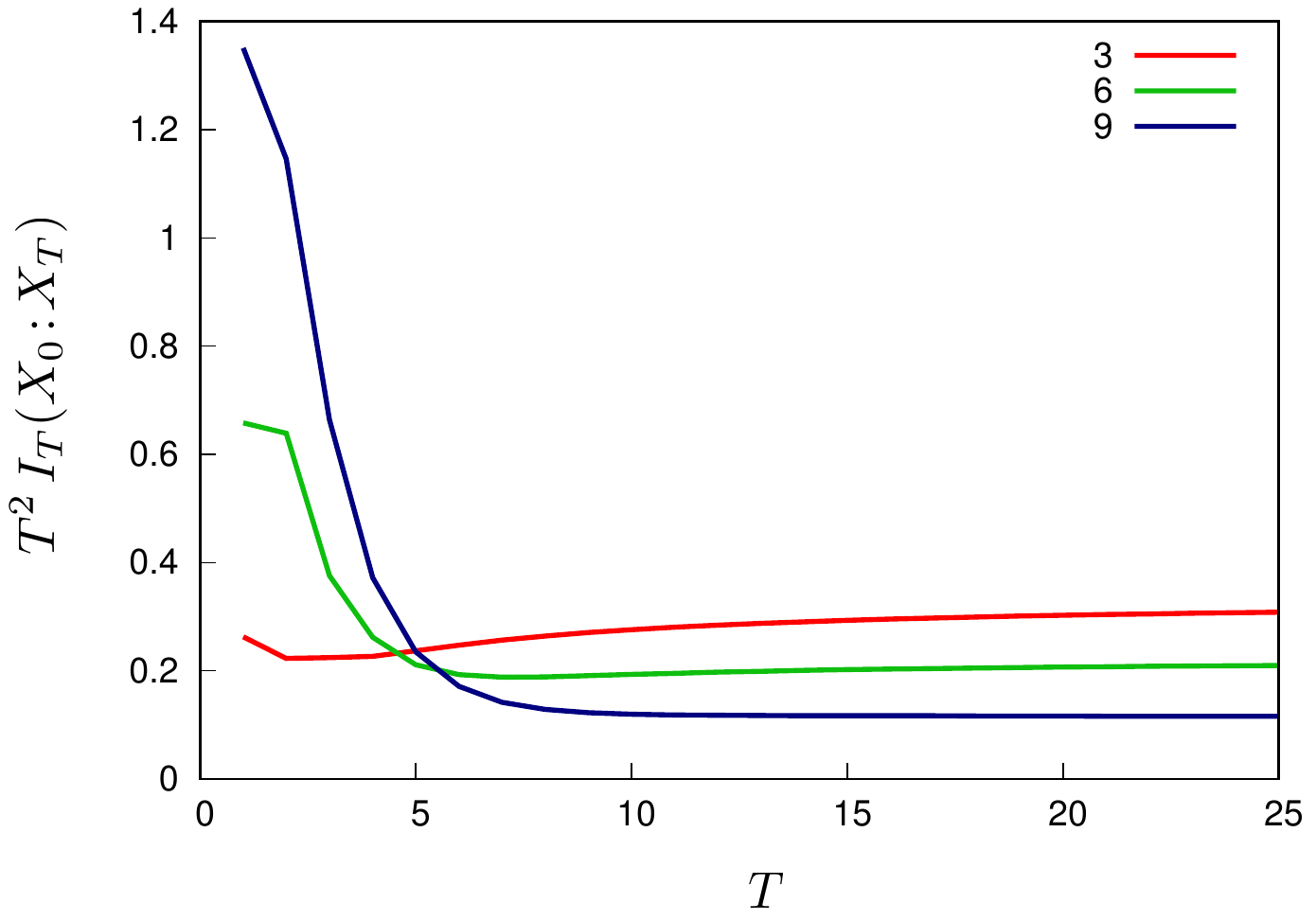}
\caption{$T^2 I_T(X_0\! :\! X_T)$, the end-to-end mutual information 
multiplied by $T^2$ for the exceptional cases of odd-$\Num$, $L\in 3\Nat$,
which have zero-energy modes.
}
\label{fig:end-to-end-mi-exceptional}  
\end{figure}

A special case is that of odd $\Num$ with $L$ a multiple of three.
Since the energy gap is zero in these cases, one anticipates a power-law
decay of the end-to-end mutual information, a very atypical behavior for
a disordered system. Precisely, the dependence is inverse square:
\begin{equation}
I_T(X_0\! :\! X_T) \sim c(L) T^{-2},
\label{eq:end-to-end-mi-zero-modes}
\end{equation}
for some constant $c(L)$, as demonstrated by the plot of 
$T^2 I_T(X_0\! :\! X_T)$ 
in Fig.~\ref{fig:end-to-end-mi-exceptional} for $L=3,6,9$. 
{\em Semi-}conservation of particle number is critical to the
explanation of this result. The zero-energy modes may be 
occupied or not, but can only step down once, at a domain wall.
A simple model keeping only the degree of freedom represented 
by the position of the domain wall (if it exists at all) suffices
to reproduce the inverse-square dependence in 
(\ref{eq:end-to-end-mi-zero-modes}). The end-to-end mutual 
information is in these cases dominated not by global fluctuations
of phase, but by the domain wall between two phases with exactly
the same entropy density.

Turning now to the planar limit $L,T\to \infty$, an important 
property of the zero-temperature TIAFM, known since the 
1960's\cite{Stephenson-64,Stephenson-70},
is that the spin-spin correlation function oscillates inside an envelope
falling off as $r^{-1/2}$ (at least along lattice directions). 
This, too, can be related to local particle number (semi-)conservation in
a very natural way.
Indeed, from the equivalence of particles to circumferential satisfied bonds,
to spin direction reversal, the spin-spin correlator along the circumferential
direction ($\hat{e}$) is translated into fermion model terms as
\begin{equation}
  \label{eq:spin-spin-to-numbers}
  \langle \sigma_i \sigma_{i+\ell{\hat{e}}} \rangle
= {\rm Re} \left\langle e^{i\pi N_\ell} \right\rangle,
\end{equation}
with $N_\ell$ the number of particles in the interval $[1,\ell]$.
It is plausible, and can be shown, that
$N_\ell$ has a Gaussian distribution for large $\ell$ Assuming that,
(\ref{eq:spin-spin-to-numbers}) immediately implies an asymptotic falloff as
an oscillatory factor times $\exp(-\pi^2\Delta N_\ell^2/2)$,
where $\Delta N_\ell^2 = \langle N_\ell^2\rangle - \langle N_\ell\rangle^2$ is
the mean-square fluctuation of number of particles in the interval.
If number fluctuations behaved in an ordinary central limiting way,
with $\Delta N_\ell^2 \sim (\ln{\ell})/{\pi^2}$, this would give 
an ordinary exponential decay. However, the zero-temperature Fermi sea
is not that compressible. A calculation reveals that
$\Delta N_\ell^2 \sim (\ln{\ell})/{\pi^2}$, asymptotically. 
Inserting that in the previous formulas results in
\begin{equation}
\langle \sigma_i \sigma_{i+\ell{\hat{e}}} \rangle \sim \ell^{-1/2} \cos n\ell,
\end{equation}
as found by Stephenson.
In the planar limit, as the earlier discussion of zero-temperature 
pure phases indicates, semi-conservation is promoted to local conservation.
This is crucial to the anomalously small fluctuations in particle number.
Without particle conservation, the ground state would inevitably exhibit
ordinary central limiting behavior of $\Delta N_\ell^2$ and therefore
exponential decay of the correlation function.


At positive temperature, all the behaviors discussed above must break down
on large enough length scales.
The mutual information has a decay rate nonvanishing in the 
$L\to\infty$ limit and the planar spin-spin correlator also decays 
exponentially.
In fermionic language, those conditions imply a nonvanishing spectral gap 
and enhanced number fluctuations, conflicting requirements in the 
presence of {\em local\/} particle conservation, indicating the
breakdown of the latter.
From a bond perspective, positive temperature brings 
thermally excited down-triangles with all bonds unsatisfied,
In string diagrams these are represented by the sixth motif 
(\includegraphics[width=18pt]{motif6}). 
With both pair creation and annihilation, the ground state $\Omega$ of
the corresponding fermion model exhibits local particle number fluctuations,
the strength of which are indicated by the expectation value
$\langle \Omega | c_{x+1}c_x \Omega\rangle$. This is a superconducting order
parameter, and can be related to the characteristics of the system in the
original statistical mechanical language as
\begin{equation}
\langle\Omega | c_{x+1} c_{x} \Omega \rangle   
=  \mathrm{Prob}\big[
{\kern -0.2em}{\includegraphics[width=30pt]{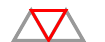}}{\kern -0.3em}\big].
\label{eq:sc-op-excited-triangles}
\end{equation}
The probability on the right is of the event that a given \downtri\ has all
three bonds unsatisfied, while its left and right neighbors have only one
unsatisfied bond (in the limit $L,T\to\infty$, the orientation data is irrelevant).

In summary,
we have shown how many characteristic properties of the triangular lattice Ising
antiferromagnet at zero temperature are elucidated by a simple equivalence 
of cylindrical TIAFM systems with fermions on a ring evolving in imaginary time.
In particular, the existence of a particle semi-conservation law in the fermionic
description turns out to play a key role. Whether other frustrated models manifest
a similar principle or whether by virtue of it the TIAFM, in addition to being an 
archetype, is a very special exemplar, remains to be seen.

\begin{acknowledgments}
This project was funded by the U.S. Department of Energy,
Office of Basic Energy Sciences, Materials Sciences and
Engineering Division under Grant No. DE-SC0010778, 
and by the National Science Foundation under
Grant No. DMR-1420620.
\end{acknowledgments}

%

\end{document}